\documentclass{revtex4-1}
\usepackage{graphicx}
\usepackage{bm}
\usepackage{amsmath,amssymb,amsfonts} 
\usepackage{mathrsfs}
\usepackage{gensymb}
\usepackage{subcaption}
\usepackage{float}
\usepackage[caption = false]{subfig}

\begin{document}

\title{Relativistic Magnetic Reconnection in the Laboratory}

\today

\author{A.~Raymond$^a$, 
C.~F.~Dong$^b$, 
A.~McKelvey$^{a,c}$, 
C.~Zulick$^{a*}$, 
N.~Alexander$^f$,
T.~Batson$^a$, 
A. Bhattacharjee$^{b}$,
P.~Campbell$^a$, 
H.~Chen$^c$, 
V.~Chvykov$^{a *\*}$,
E.~Del Rio$^f$,
P.~Fitzsimmons$^f$,
W.~Fox$^b$, 
B.~Hou$^{a}$,
A.~Maksimchuk$^a$, 
C.~Mileham$^d$, 
J.~Nees$^a$,
P.~M.~Nilson$^d$, 
C.~Stoeckl$^d$, 
A.~G.~R.~Thomas$^{a,e,g}$, 
M.~S.~Wei$^f$, 
V.~Yanovsky$^{a}$, 
L.~Willingale$^{a,e,g}$, 
K.~Krushelnick$^a$}

\affiliation{$^a$ Center for Ultrafast Optical Science, University of Michigan, Ann Arbor, Michigan 48109, USA.\\
$^b$ Princeton Plasma Physics Laboratory, Princeton University, Princeton, NJ 08540.\\
$^c$ Lawrence Livermore National Laboratory, 7000 East Ave, Livermore, CA 94550.\\
$^d$ Laboratory for Laser Energetics, Rochester University, Rochester, NY 14623.\\
$^e$ Physics Department, Lancaster University, Lancaster LA1 4YB, United Kingdom.\\
$^f$ General Atomics, 3550 General Atomics Court San Diego, CA 92121-1122.\\
$^{g}$ The Cockcroft Institute of Accelerator Science and Technology, Sci-Tech Daresbury, Keckwick Lane, Daresbury, Warrington WA4 4AD, United Kingdom.\\
$^*$ Current address: Plasma Physics Division, U.S. Naval Research Laboratory, Washington, DC 20375.\\
$^{\*\*}$ Current address: ELI-Alps, Szeged, Pint\'er J\'ozsef Utca, 6728 Hungary. }

\begin{abstract}
Magnetic reconnection is a fundamental plasma process involving an exchange of magnetic energy to plasma kinetic energy through changes in the magnetic field topology.
In many astrophysical plasmas magnetic reconnection plays a key role in the release of large amounts of energy \cite{hoshino1}, although making direct measurements is challenging in the case of high-energy astrophysical systems such as pulsar wind emissions \cite{lyubarsky1}, gamma-ray bursts \cite{thompson1}, and jets from active galactic nuclei \cite{liu1}.
Therefore, laboratory studies of magnetic reconnection provide an important platform for testing theories and characterising different regimes.
Here we present experimental measurements as well as numerical modeling of relativistic magnetic reconnection driven by short-pulse, high-intensity lasers that produce relativistic plasma along with extremely strong magnetic fields.
Evidence of magnetic reconnection was identified by the plasma's X-ray emission patterns, changes to the electron energy spectrum, and by measuring the time over which reconnection occurs.
Accessing these relativistic conditions in the laboratory allows for further investigation that may provide insight into unresolved areas in space and astro-physics.
 \end{abstract} 

\maketitle
Laboratory plasmas can be used to investigate magnetic reconnection within magnetic fusion energy devices such as Tokamaks \cite{Tokamak}, using dedicated experiments such as the Magnetic Reconnection Experiment (MRX) \cite{Yamada_PoP_1997}, or using laser-driven plasmas \cite{Nilson_PRL_2006, Li_PRL_2007, Nilson_PoP_2008, Willingale_PoP_2010, Zhong_NP_2011, Dong_PRL_2012, Fiksel_PRL_2014, Rosenberg_PRL_2015, Rosenberg_NC_2015}.
This enables the study of fundamental energy transfer processes occurring during magnetic field rearrangement over a varying range of plasma parameters. Different regimes can therefore be examined and experiments may provide complimentary data.

Previous laser-driven magnetic reconnection experiments used kilojoule class, nanosecond duration laser pulses (long-pulse regime) focused to moderate intensities, $I \simeq 10^{14-15}$ Wcm$^{-2}$, to heat a solid target and create two neighbouring plasma bubbles \cite{Nilson_PRL_2006, Li_PRL_2007, Nilson_PoP_2008, Willingale_PoP_2010, Zhong_NP_2011, Dong_PRL_2012, Fiksel_PRL_2014, Rosenberg_PRL_2015, Rosenberg_NC_2015}.
Azimuthal megagauss magnetic fields in each expanding plasma are driven together by frozen-in-flow, i.e.\ the bulk motion of the plasma, or hot electron flows \cite{Willingale2010,Joglekar2014}.
This experimental regime is characterised by a high Lundquist number and large system size compared to the electron and ion inertial lengths, justifying dimensionless parameter scalings with many astrophysical systems \cite{Ryutov_AJSS_2000}.


Recent particle-in-cell (PIC) simulations of magnetic reconnection in the long-pulse regime \cite{Totorica_PRL_2016} demonstrate that electrons may be accelerated by the reconnection fields to energies exceeding an order of magnitude larger than the thermal energy.
Additional PIC simulations in the short-pulse regime have shown strong reconnection features during in near-critical and low density environments \cite{Ping_PRE_2014, Gu_PRE_2016}. 

However, until now, the extremely energetic class of astrophysical phenomena including high-energy pulsar winds, gamma ray bursts, and jets from galactic nuclei \cite{lyubarsky1,Tokamak,thompson1,liu1}, where the energy density of the reconnecting fields exceeds the rest mass energy density ($\sigma \equiv B^2/(\mu_0 n_e m_e c^2)>1$), have been inaccessible in the laboratory. This regime defines relativistic reconnection.

Here we present the use of much higher intensity laser pulses ($I > 10^{18}$ Wcm$^{-2}$), that generate a dense, relativistic electron plasma within the focal volume during the interaction with a solid foil target. In this regime, magnetic field generation mechanisms and transport are governed by the relativistic electron population and its dynamics.
The isotropic expansion of the electrons rapidly sets up a space-charge field at the target-vacuum interface, effectively confining the hot electrons to expand radially along the target surface \cite{Schumaker_PRL_2013}.
These currents generate an azimuthal magnetic field with $\mathscr{O}($100 MG) magnitude expanding radially at $v_B \sim c$ \cite{Schumaker_PRL_2013}.
Focusing two of these laser pulses in close proximity on the target surface produces a reconnection geometry with plasma characteristics of the relativistic reconnection regime.

Here, experimental data is compared with three-dimensional PIC modeling to present evidence for magnetic reconnection in the relativistic regime with parameters based on the experimental conditions.
Experimentally, an x-ray (copper K$_{\alpha}$) imaging technique visualised the fast electrons accelerated in the reconnection region to provide spatial information about this region's extent.
This x-ray signal also enables a temporal duration measurement of the region's lifetime.
Simulations show the plasma density and magnetic field characteristics in the reconnection region satisfy $\sigma>1$, indicating these experiments are in the relativistic reconnection regime.

\begin{figure}[h!]
\begin{center}
\includegraphics[width=1\textwidth]{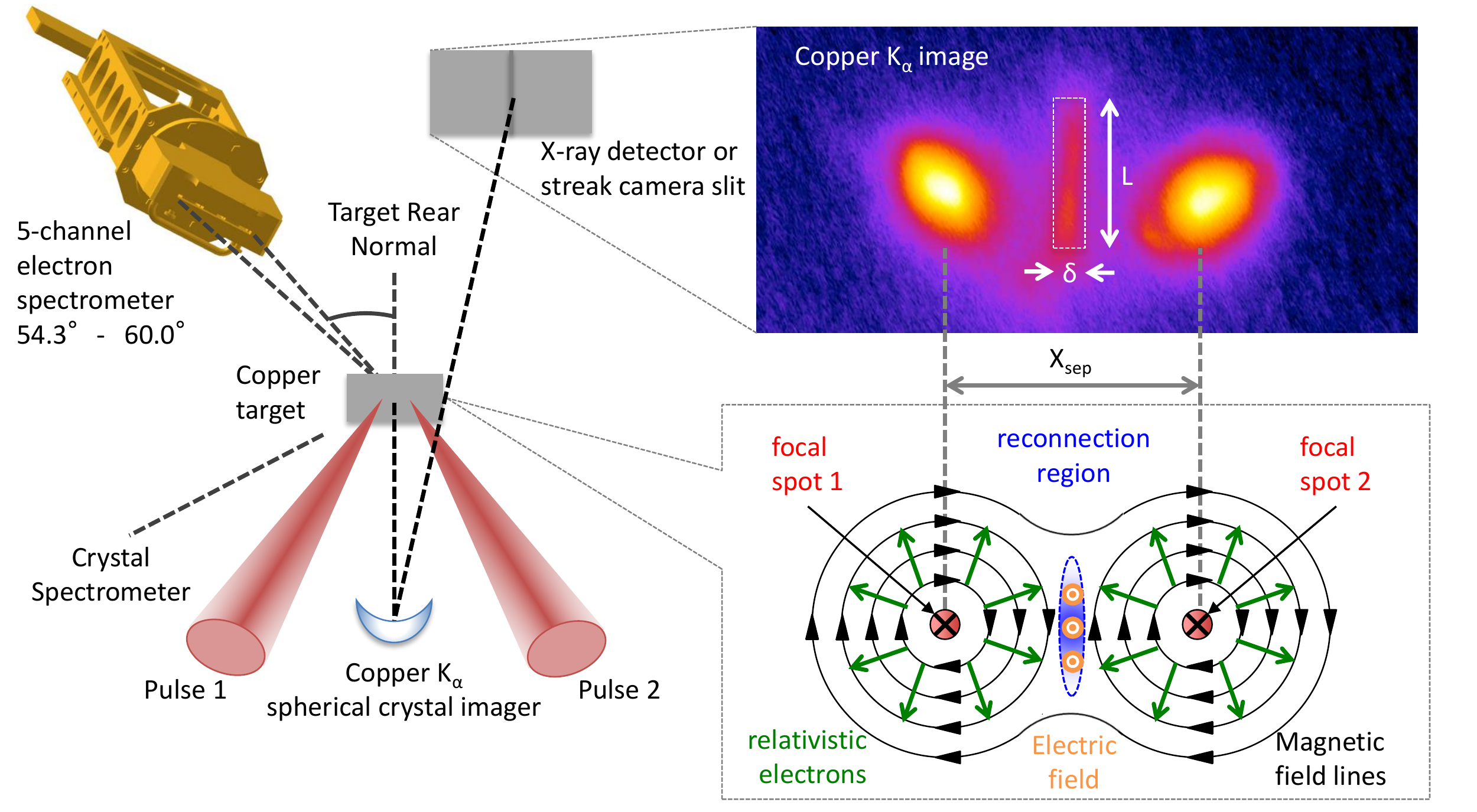}
\caption{A schematic of the experimental geometry. The spherical copper K$_{\alpha}$ crystal images the front side of the target onto a detector. A typical K$_{\alpha}$ image is shown with the reconnection layer highlighted in the dashed box with of length ($L$) and width ($\delta$) labeled. A physical picture of the interaction illustrates the two azimuthal magnetic fields expanding into the reconnection region where a target normal electric field accelerates the electrons into the dense target to generate the copper K$_{\alpha}$ emission in the midplane.
\label{fig1}}
\end{center}
\end{figure}

The experiments focused dual short-pulses onto $\mu$m-scale copper foil targets to two focal spots separated by a distance $X_{sep}$.
A generalised experimental schematic and diagram of the two-spot field geometry with corresponding magnetic and electric fields is depicted in Figure \ref{fig1}.
When the anti-parallel magnetic fields meet in the midplane between the interaction site, the field lines break and reconnect within a reconnection layer, deflecting inflowing electrons and supporting a target-normal electric field supported by the Hall effect and by thermal pressure. This localised electric field generates a current sheet, with electrons being accelerated into the dense regions of the plasma. These fast electrons undergo ionising collisions with K-shell electrons as they propagate into the dense target. K-alpha X-ray emission occurs as the L-shell electrons transition to the K-shell. Therefore, by imaging this K-alpha emission, we are able to produce a map of the current sheet generated between the magnetic field regions and hence diagnose the reconnection process.

A  spherically bent X-ray crystal \cite{kalphacrystalref} was used to image the copper $K_\alpha$ emission (8.048 keV) from the target front-side as L-shell electrons transition to the K-shell.

\begin{figure}[h!]
\begin{center}
\includegraphics[width=1\textwidth]{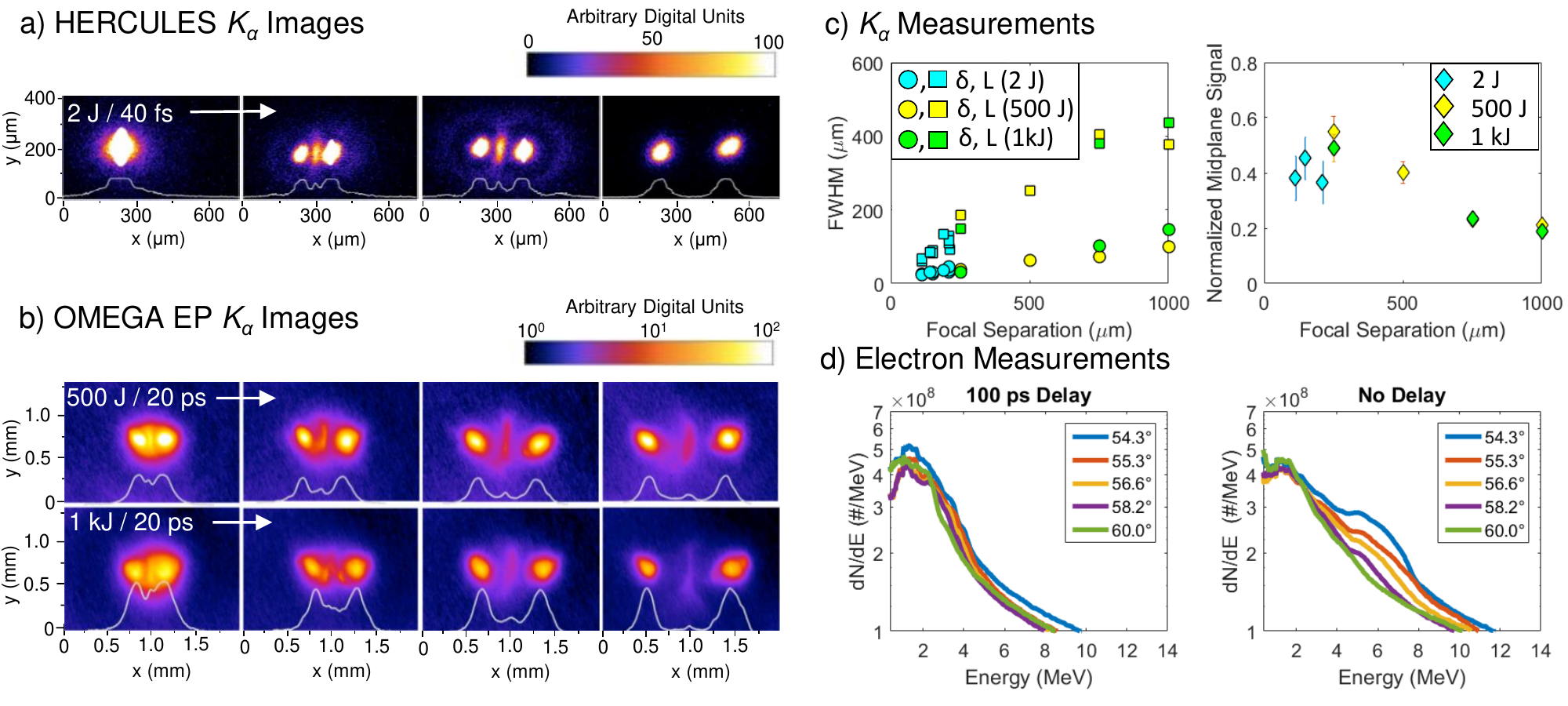}
\caption{Front-side copper K${_\alpha}$ images from focal spot separation scans using (a) the \sc{Hercules}\rm{} laser and (b) the \sc{OMEGA EP}\rm{} laser. 50 $\mu$m horizontal line-outs are superimposed.
c) The enhanced midplane signal F.W.H.M. dimensions (left) and integrated signal (right) normalised to the per-shot average of the integrated signal density from the focal spot regions.
d) The electron spectra from the \sc{OMEGA EP}\rm{} multi-channel spectrometer at angles from the transmitted laser axis in the case of a 100 ps pulse-to-pulse delay (no reconnection expected) and no pulse-to-pulse delay (reconnection expected). Angles are given with respect to the rear target normal.
}
\label{fig2}
\end{center}
\end{figure}

With this experimental geometry, separation scans of the focal spots were performed with the \sc{Hercules}\rm{} laser at the University of Michigan (2 J, 40 fs pulses focused to extreme intensities $2 \times 10^{19} \; \rm{Wcm}^{-2}$) and \sc{OMEGA EP}\rm{} laser at the Laboratory for Laser Energetics (500 J / 1000 J, 20 ps focused to comparable intensities of $(1.2 / 2.5) \times 10^{18} \; \rm{Wcm}^{-2}$).
Two bright $K_\alpha$ sources corresponding to the target heating within the focal volume were observed (Figure \ref{fig2} (a),(b)).
Additionally a separation-dependent enhancement of the $K_{\alpha}$ radiation in the midplane between the two focal spots was measured corresponding to the electrons accelerated within the reconnection layer into the target.
From \sc{Hercules}\rm{}, the midplane $K_{\alpha}$ enhancement was observed between the foci until a cutoff of $X_{sep}>215 \; \mu$m, at which point the $1/r$ reduction of the radially expanding magnetic field’s magnitude suppresses reconnection. 
The higher-energy \sc{OMEGA EP}\rm{} produced similar K$_{\alpha}$ images reflecting the separation-dependence of the reconnection process.

To quantitatively compare the $K_\alpha$ emission characteristics, the full-width-at-half-maximum dimensions of the reconnection region are plotted versus $X_{sep}$, as well as the integrated midplane signal in Figure \ref{fig2} (c).
Linear trends were observed for both the length $L$ and width $\delta$ of the reconnection region versus focal separation suggesting the regime accessed at either facility are comparable, despite the drastic change in pulse parameters.
The peak intensities used were comparable and the generation of electrons has a strong intensity dependence, suggesting that the defining characteristic is the relativistic flow of electrons into the reconnection layer.
The dependence of the reconnection layer dimensions upon the spot separation can be expressed as $L(\mu \rm{m}) \sim 0.5 \times X_{sep}(\mu \rm{m})$ and $\delta(\mu \rm{m}) \sim 0.15 \times X_{sep}(\mu \rm{m})$; these fit the data with a coefficient of determination $R^2 > 0.7$ for experiments using both \sc{Hercules}\rm{} and \sc{OMEGA EP}\rm{}. The aspect ratio of the reconnection layer (in this regime identified as the rate at which flux conservation is violated in the reconnection process) is $\delta/L \approx 0.3$, suggesting a rapid conversion of magnetic potential energy to plasma energy.


These trends demonstrate the target-normal electric field generated during reconnection increases in magnitude as $X_{sep}$ decreases.
This corresponds to a higher magnetic field magnitude near the midplane and therefore produced a larger in-plane electron current inducing the midplane signal.
The reconnection layer signal width $\delta$ increases with focal spot separation, suggesting that there is an increased spread of target-directed electron trajectories given the presumably steeper density gradient (less pre-plasma) within the region.
Further, the increasing length of the reconnection layer $L$ with $X_{sep}$ suggests a consistent geometric
dependence upon the radius of curvature of the magnetic field lines.

Integrating the mid-plane signal (Figure \ref{fig2} (c)) determines the relative efficiency of mid-plane current generation. Both data sets from \sc{OMEGA EP}\rm{} and \sc{Hercules}\rm{} suggest that smaller separations yield stronger reconnection features, until the point when reconnection is suppressed (as seen in the latter) due to a more sensitive dependency upon slight beam mis-timings and target deformation.


Comparisons of front-side to back-side $K_\alpha$ measurements conducted on \sc{OMEGA EP}\rm{} allowed the estimation that the $K_\alpha$ photons emanate from depths of $> 10$ $\mu$m from the front-side (corresponding to electron energies $\sim$100 keV).
Measuring the $K_\beta/K_\alpha$ ratio from the midplane region with a von Hamos crystal spectrometer \cite{vonHamosCrystalSpectrometer} suggested the K-photons originated from a cold (and correspondingly high-density) region of the plasma \cite{nilson_kalphakbeta}, further indicating acceleration of electrons into the plasma gradient by a strong electric field set up during reconnection. 
X-ray pinhole camera images found no midplane emission between $2-6$ keV, ruling out collisional heating between the two plasmas as a source of the $K_\alpha$ enhancement.

The relatively long pulse duration utilised on \sc{OMEGA EP}\rm{} (20 ps) enabled time resolved X-ray streak measurements of the mid-plane $K_\alpha$ emission. Specialised targets comprised of copper only in the midplane were irradiated, resulting in measured durations of 25$\pm 5$ ps (comparable to the burst of relativistic electrons). Such a short duration indicates a significantly fast conversion of magnetic potential energy to relativistic thermal energy (much faster than the $\mathscr{O}($1 ns) hydrodynamic collision time of the plasma).

To search for direct signatures of electrons accelerated by the reconnection electric field, a 5-channel electron spectrometer was utilised on \sc{OMEGA EP}\rm{} at the target-rear (Figure \ref{fig2} (d)).
It showed angularly dependent non-thermal features superimposed upon a quasi-Maxwellian energy distribution for the electrons, as would be expected if a population of directionally accelerated electrons propagated through the target and escaped into vacuum.
When a 100 picosecond delay was introduced between the pulses, the nonthermal portion of the spectrum was suppressed, supporting their association with reconnection.

Therefore the experimental results infer strong evidence for magnetic reconnection from several perspectives: a consistent, localised enhancement of $K_\alpha$ emission from the focal midplane; a short duration of this emission; evidence that the K-photons originate from deep within the target; and electron spectra consistent with nonthermal acceleration of electrons in the midplane region.

A fully relativistic, three-dimensional simulation of the reconnection geometry most closely resembling the \sc{Hercules}\rm{} short-pulse scenario was conducted using the PIC code OSIRIS \cite{Fonseca_ICCS_2002}.
The laser pulse had a 40 fs full-width-at-half-maximum pulse duration and  focused intensity of $2 \times 10^{19}$ Wcm$^{-2}$.
The pulse was normally incident upon an electron plasma with $n_{max}=30n_{crit}$ (where $n_{crit}=\frac{\epsilon_0 m_e}{e^2}\omega_{0}^2$ is the critical plasma density) and preplasma scale length $l=\lambda=810$ nm, and with stationary ions.
Periodic boundaries in the transverse directions were utilised to result in an effective spot-to-spot separation of $388$ $\omega_0/c$ (50 $\mu$m).

\begin{figure}[h!]
\begin{center}
\includegraphics[width=1\textwidth]{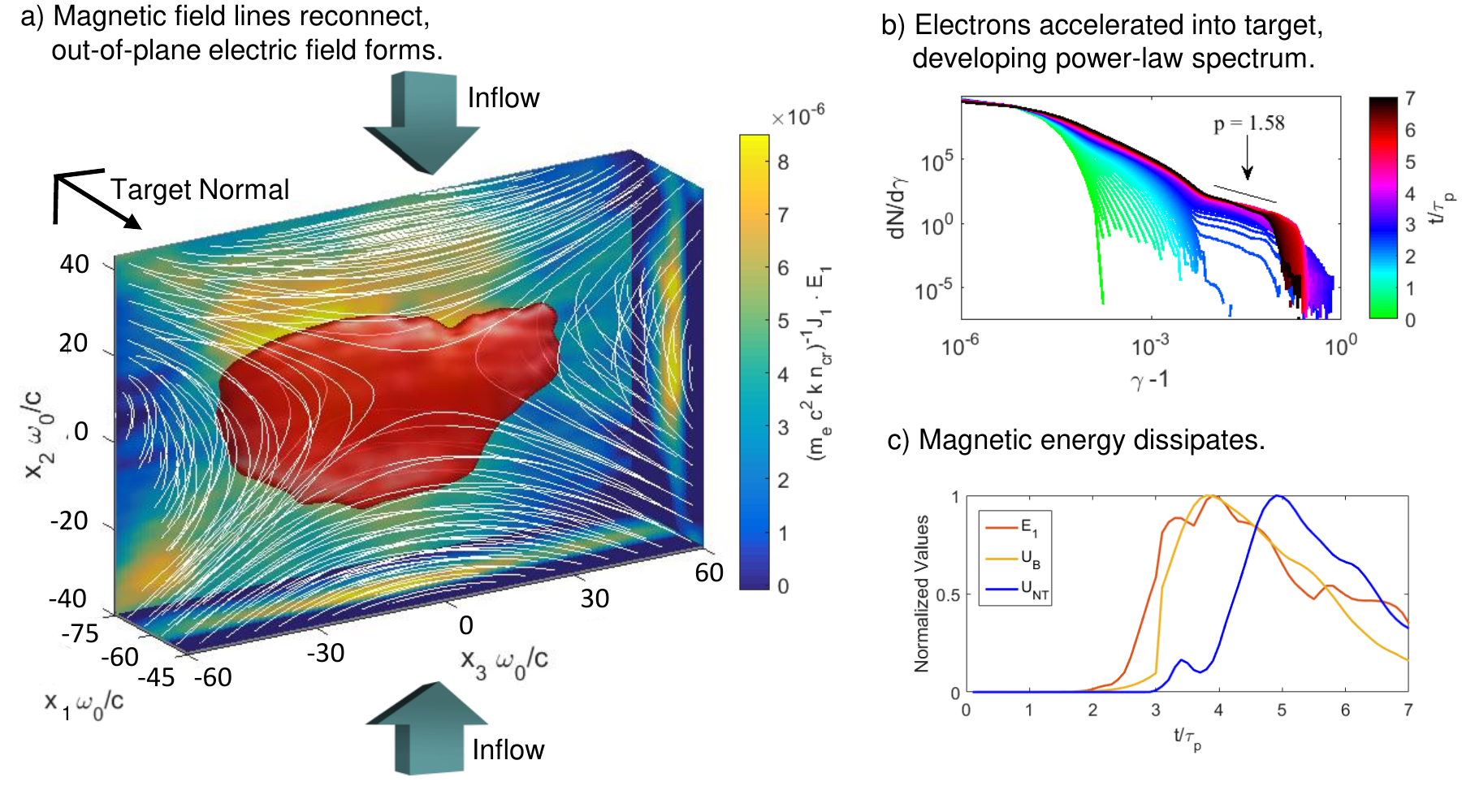}
\caption{a) A 3-D graphic of magnetic field streamlines (white), the associated reconnection electric field (displayed as a red iso-surface with magnitude 95 gigavolts/m), and the value of $E_1 \cdot J_1$ evaluated in center cuts through the displayed volume (box faces). The interaction sites are located 194 $\omega_0/c$ away along $x_2$, and $x_1$ is in the target-normal direction. b) The temporal evolution of the electron $\gamma=\sqrt{1+(p_e/m_ec)^2}$ spectrum within the midplane region, acquiring a hard power law spectrum when reconnection begins. c) The temporal behavior of the maximum reconnection electric field magnitude ($E_1$), energy in nonthermal electrons ($U_{NT}$), and magnetic potential energy ($U_B$) evaluated in the reconnection region (all quantities are normalized to their respective maximum values). The simulation pulse duration is $\tau_p=40$ fs.}
\label{fig3}
\end{center}
\end{figure}

An isotropic, Maxwellian population of electrons with $\sim$1 MeV temperature was seen to be generated from the interaction site(s) after irradiation, followed by confinement of the fast electrons along the plasma surface.
The maximum azimuthal B-field magnitude within the interaction site is 35 MG.
The counter-streaming surface electrons and their associated azimuthal magnetic fields reach the midplane within $2$ pulse durations, when rapid reconnection of the magnetic field lines within a region of $\delta \times L = 50 \times 200 \omega_{0}/c$ was observed (a dimensional ratio comparable to the experimental measurements).

As anti-parallel magnetic field lines converged and began to reconnect, an out-of-plane electric field with $E_{peak}/cB_{R0}\approx 0.3$ formed, where $B_{R0} \approx 2$ MG is the azimuthal magnetic field magnitude in the vicinity of the reconnection layer.
In addition, an out-of-plane quadrupole magnetic field pattern develops indicating Hall-like reconnection \cite{Uzdensky_PoP_2006} (shown in Figure \ref{hall_fig}). Evaluating $E \cdot J$ in the midplane region shows a localization of work done on electrons in the target normal direction (Figure \ref{fig3} (a)).
Electrons are accelerated into the target, developing a nonthermal spectral component in addition to a Maxwellian low-energy portion (Figure \ref{fig3} (b)).
The nonthermal component may be fitted by a power law $dN/d\gamma \propto \gamma^{-1.6}$, which is consistent for relativistic reconnection \cite{relrec_pl_cite}.
The maximum value of $\gamma$ was found to be consistent with the work done by the peak electric field over the reconnection time.

Figure \ref{fig3} (c) depicts the temporal evolution of three simulation variables: the total magnetic energy density within the vicinity of the reconnection region, the maximum value of the target-normal electric field within this same region (with the capacitive sheath-field removed), and the energy in the nonthermal portion of the spectra presented in Figure \ref{fig3} (b). 
Significant energy is imparted to the nonthermal electrons as the magnetic energy dissipates, which is consistent with reconnection.
The temporal F.W.H.M. of the nonthermal electron energy trace is $\approx 2$ pulse durations, demonstrating the rapid dissipation of magnetic energy as suggested by the experimental data.

The interaction is characterised by a plasma beta of $\beta_e = P_{plasma}/P_{magnetic}\sim 50$ and electron skin depth $c/\omega_{pe} = 2.4$ $\mu$m $< l$ (the current sheet length), meeting the criterion for Hall-like reconnection.
96 \% of the electron gyroradii $r_g=m_ev_{\perp}/|q|B$ are $\leq \delta$.
The parameter $\sigma$ exceeds 10 within the reconnection region, confirming the relativistic reconnection regime is accessed.

The characteristics of the reconnection observed in the simulation and those measured experimentally both indicate an efficient transfer of magnetic potential energy to non-thermal energies over incredibly short timescales, within in a regime characterised by the involvement of relativistic particles.
Further, the rate of reconnection as measured experimentally by the reconnection layer aspect ratio match with that deduced from the simulation's electric field characteristics.

The measurements reported here represent the first experimental generation of magnetic reconnection in the regime of relativistic plasma inflows and potentially high values of $\sigma$.
The use of two relativistically intense laser pulses may thus be a prime test-bed for conducting reconnection experiments relevant to high-energy astrophysical processes, for example by measuring high-energy photons to compare to those measured from gamma-ray bursts. Access to regimes of even higher magnetic fields may be made available by next-generation Petawatt laser facilities such as ELI \cite{Charalambisdis_CLEO_2013}, leading to a method to characterise the reconnection in relativistic electron-positron plasmas in the laboratory.

\section*{Supplementary Information: Methods}

The \sc{Hercules}\rm{} laser facility at the University of Michigan is a Ti:Sapphire system with $\lambda=810$ nm and F.W.H.M. pulse durations of 40 fs delivered 100 TW pulses to the target chamber. The amplified spontaneous emission (ASE) contrast was reduced with the Cross-Wave Polarisation (XPW) \cite{Chvykov_OL_2006} correction technique to $10^{-11}$. An f/3 paraboloidal mirror was split along its symmetry axis with one half vertically motorised to allow two wavefront-corrected laser foci to be adjustably separated with $25$ $\mu$m precision. One focal spot had an approximate F.W.H.M. area of 174 $\mu$m$^2$ and 37.5 TW, while the other had an approximate F.W.H.M. area of 337 $\mu$m$^2$ and 63 TW. Therefore both spots had approximately equal focused intensities of 2$\times 10^{19}$ Wcm$^{-2}$.



A spherically bent, imaging X-ray crystal (Quartz 211) with a $2d=3.082 \textrm{\r{A}}$ lattice spacing was positioned 242 mm from the target and created an image at 1260 mm, yielding a magnification of 5.2 onto an Andor iKon-M CCD. The CCD had 13 $\mu$m pixels and at a quantum efficiency of 45\% for copper $K_{\alpha}$ light and the spatial resolution of the image was 15 $\mu$m. 

The \sc{OMEGA EP}\rm{} facility provided independent laser pulses with a F.W.H.M. duration of 20 ps and energies of either 500 J or 1000 J. The focal spot F.W.H.M. was 13$\pm2$ $\mu$m. Pulse arrival on target was measured within $\pm 5$ ps by utilising an ultrafast X-ray streak camera.
 
A similar spherically bent imaging X-ray crystal diagnostic was used on \sc{OMEGA EP}\rm{}, comprised of Quartz with a 50 cm radius of curvature, $2d=3.082$ \r{A} lattice spacing, and 2131 Miller indices. A 50 $\mu$m alumised mylar blast shield was protected the crystal from damage. The crystal imaged with a magnification of $\approx 10$ onto an image plate (IP) detector positioned in the opposing diagnostic port, satisfying the Bragg condition of 1.31$^{\circ}$. The IP was shielded by lead to minimise X-ray background, and the line-of-sight was blocked by a cylinder of beryllium. The final spatial resolution was 10 $\mu$m. Depending on the experiment, the X-ray crystal viewed either the front or the rear of the target surface, slightly off-axis.

X-ray pinhole cameras (XRPHC) were employed on \sc{OMEGA EP}\rm{}. The detectors were charge injection devices (CID) filtered to detect photons between 2-6 keV.

\begin{figure}[h!]
\begin{center}
\includegraphics[width=.6\textwidth]{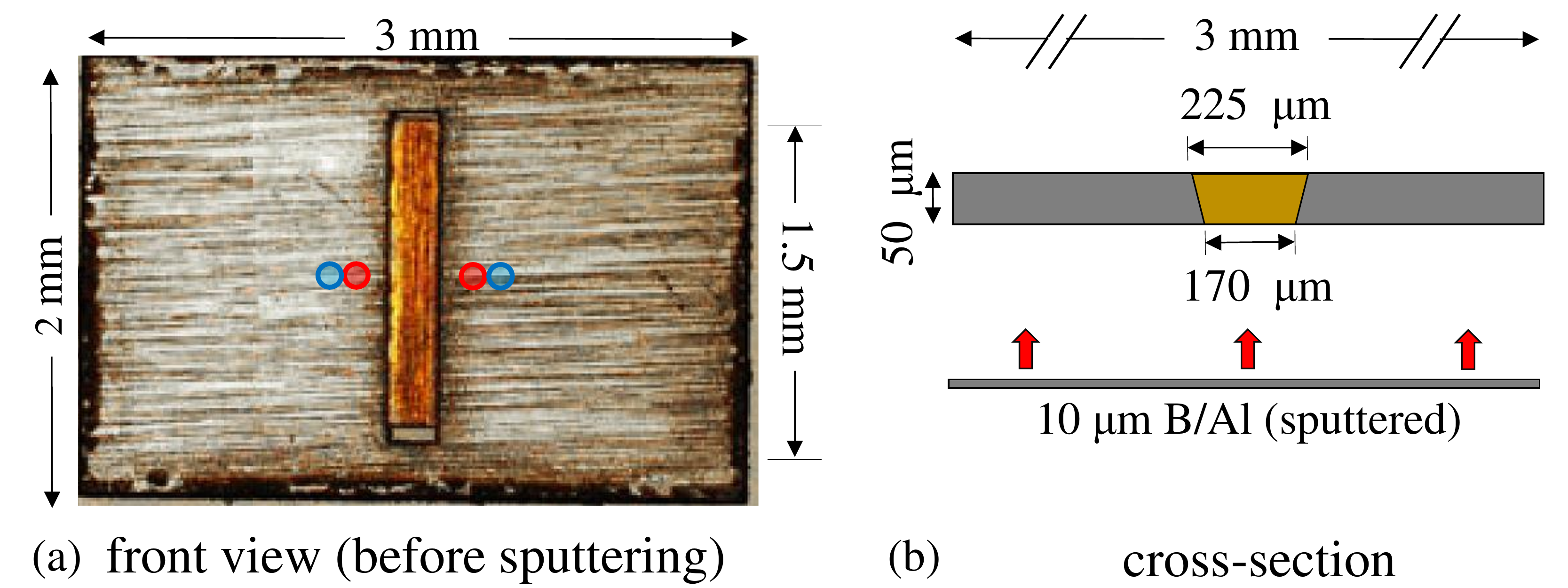}
\caption{The target design used to make the OMEGA EP temporal duration of the midplane emission measurements. a) A front-view image of the target (before sputtering); the circles indicate the focal spot positions corresponding to a 500 $\mu$m separation (red) and 750 $\mu$m separation (blue). While the circles are displayed large for visibility, the focal radius at which 80\% of the laser energy was contained equaled 18 $\mu$m on average. b) The target's cross section, demonstrating the boron/aluminium layer sputtered over the target's front side.}
\label{target_geo_manuf}
\end{center}
\end{figure}

The targets for the \sc{Hercules}\rm{} shots were 12 $\mu$m thick copper foils. For the separation scan performed on \sc{OMEGA EP}\rm{}, 50 $\mu$m thick copper foils were used. The targets used to determine the duration of the $K_\alpha$ emission were manufactured by General Atomics. The schematic in Figure \ref{target_geo_manuf} demonstrates the target geometry. All \sc{OMEGA EP}\rm{} targets were mounted to 50 $\mu$m silicon dioxide stalks.

To determine the temporal duration of the midplane $K_\alpha$ emission, specially designed targets (as shown in Figure \ref{target_geo_manuf}) were irradiated by \sc{OMEGA EP}\rm{}. The spherically bent X-ray crystal collected and imaged the emitted copper $K_\alpha$ radiation onto the slit of an utlrafast X-ray streak camera. To ensure that the radially expanding surface electrons were not impeded by resistive magnetic fields due to steep resistivity gradients between target materials \cite{Wei_PoP}, a 10 $\mu$m layer of B/Al was sputtered onto the front surface of the target (upon which the lasers are incident). The X-ray crystal imaged the Cu bar onto the slit of a picosecond streak camera (3000 \AA \textbf{ } CsI photocathode and 600 ps sweep speed) with 6$\times$ magnification.

A multi-channel electron spectrometer developed by Osaka University was utilised to detect angularly dependent electron spectra. It was positioned such that an angle of $0^\circ$ corresponds to the (transmitted) laser axis.

The PIC code OSIRIS was utilised to run a 3-D simulation of the fast reconnection geometry most closely resembling the \sc{Hercules}\rm{} short-pulse scenario. 
The simulation box had dimensions $X_2 \times X_3=387\omega_0/c \times 775.8$ $\omega_0/c$ in the laser transverse direction and  $X_1=185$ $\omega_0/c$ in the laser propagation axis, with a resolution of 6.3 cells per $c/\omega_0$ (40 cells per $\lambda$).
The plasma was an electron plasma with $n_{max}=30$ $n_{crit}$, $3 \times 3 \times 3$ particles per cell, and stationary ions.

The laser pulse (40 fs F.W.H.M. and $a_0$=3 traveling along $X_1$) was normally incident upon a plasma with $n_{max}=30n_{crit}$ and preplasma scale length $L=\lambda$, at a point one third of the way through $X_2$. Periodic boundaries in $X_2$ were utilised to result in an effective spot-to-spot separation of $X_{sep}=776$ $\omega_0/c$ (50 $\mu$m), as well as in $X_3$ to avoid artificial boundary effects. Thermal boundaries were utilised in $X_1$ for the particles to prevent electron refluxing.

\begin{figure}[h!]
\begin{center}
\includegraphics[width=.6\textwidth]{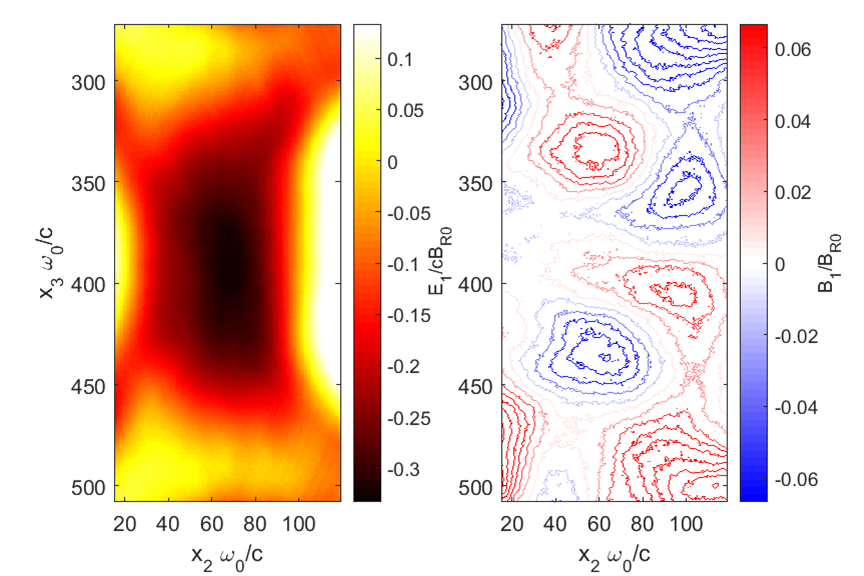}
\caption{(a) The target-normal electric field with the capacitive sheath field subtracted, and normalized to $cB_{R0}$. (b) The magnetic field in the target-normal direction within the same region normalized to $B_{R0}$, demonstrating the quadrupole magntic field pattern indicative of Hall-like reconnection.}
\label{hall_fig}
\end{center}
\end{figure}

Figure \ref{hall_fig} demonstrates the target-normal electric field in the reconnection region and the out-of-plane quadrupole magnetic field configuration in the vicinity of the reconnecting X-point, indicating a significant contribution of the Hall term in the generalized Ohm's law to the reconnection electric field. In this plot (and in the references made to the reconnection electric field in the text), the capacitive sheath field across the entire plasma surface was approximately removed from the reconnection region by subtracting the value of $0.5[max(E_{T1})+min(E_{T1})]$ (where $E_{T1}$ is the total electric field) evaluated within the same region.

%
%

\section{Acknowledgments}
This material is based upon work partially supported by the Department of Energy National Nuclear Security Administration under Award Number DE-NA0002727. The authors gratefully acknowledge technical assistance from the Laboratory for Laser Energetics, and appreciate the use of NASA's High End Computing Capability. We also thank the OSIRIS consortium for developing and maintaining their state-of-the-art PIC code, citing grant number NSF ACI 1339893. HC is supported by the DOE (Grant No. DE-FG52-09NA29041). 

\section{Contributions} 
A.R., A.McK. and C.Z. prepared and conducted the HERCULES experiments under the supervision of A.Mak., A.G.R.T., L.W. and K.K., with experimental support from V.C., B.H., J.N. and V.Y.. A.R. designed and conducted the OMEGA EP experiments with the assistance of L.W., K.K., and diagnostic support and analysis was provided by C.M., P.M.N., C.S. (copper K$_{\alpha}$ imaging and streak camera), M.S.W. (x-ray spectrometer) and H.C. (electron spectrometer). OMEGA EP target development and fabrication was performed by the General Atomics Inertial Fusion Technology group, with the assistance of P.F., N. A. and E.D.R.. Data analysis was performed by A.R. with assistance from C.Z., P.C., M.S.W., H.C., P.M.N., and C.S.. The simulations were performed and analysed by A.R. and C.F.D. with discussions with A.G.R.T., L.W., W.F. and K.K.. The manuscript was written and prepared by A.R., along with L.W. and K.K..


\end{document}